\documentclass[aps, prd, reprint, preprintnumbers, showpacs, floatfix, nofootinbib, superscriptaddress,
nofootinbib]{revtex4-2}
\usepackage[utf8]{inputenc}
\usepackage{amssymb}
\usepackage{hhline}
\usepackage{amsmath}
\usepackage{mathtools}
\usepackage[dvipsnames]{xcolor}
\usepackage{multirow,tabularx}
\usepackage{natbib}
\usepackage{orcidlink}
\usepackage{natbib}
\usepackage{xspace}
\usepackage{xstring}
\usepackage{parskip}
\allowdisplaybreaks
\usepackage{braket}

\newcommand{\mnras}{Mon. Not. R. Astron. Soc.}

\newcommand{\gsim}{\raisebox{-0.13cm}{~\shortstack{$>$ \\[-0.07cm]
      $\sim$}}~}

\usepackage{hyperref}
\hypersetup{%
     colorlinks = true,%
     linkcolor = Blue,%
     citecolor = Blue,%
     filecolor = Blue,%
     urlcolor = Blue%
     }%

\definecolor{tabblue}{RGB}{31, 119, 180}
\definecolor{darkblue}{RGB}{0, 0, 120}
\definecolor{tabred}{RGB}{214, 39, 40}
\definecolor{tabgreen}{RGB}{44, 160, 44}
\definecolor{tabgray}{RGB}{100, 100, 100}
\begin{document}

\title{The Sensitivity of PUEO to Cosmogenic Neutrinos and Exotic Physics Scenarios}

\author{Angelina Sherman\orcidlink{0009-0007-2369-258X}}
\affiliation{Department of Physics, Wisconsin IceCube Particle Astrophysics Center, University of Wisconsin, Madison, WI, 53706}

\author{Ke Fang\orcidlink{0000-0002-5387-8138}}
\affiliation{Department of Physics, Wisconsin IceCube Particle Astrophysics Center, University of Wisconsin, Madison, WI, 53706}

\author{Dan Hooper\orcidlink{0009-0004-2456-1221}}
\affiliation{Department of Physics, Wisconsin IceCube Particle Astrophysics Center, University of Wisconsin, Madison, WI, 53706}

\date{\today}

\begin{abstract}

Several observatories designed to detect ultrahigh-energy neutrinos are planned for the next decade. The most imminent of these is the Payload for Ultrahigh Energy Observations (PUEO), a long-duration balloon-based experiment that will provide unprecedented sensitivity to neutrinos with energies in the range of $\sim 1 - 1000 \, {\rm EeV}$. In this work, we assess the scientific reach of PUEO. In particular, we evaluate the sensitivity of this observatory to cosmogenic neutrinos and, in turn, to the proton fraction of the ultrahigh-energy cosmic-ray spectrum. We also consider the potential of PUEO to probe scenarios in which neutrinos are produced through the decays of ultraheavy dark matter particles or are radiated from cosmic strings. We find that PUEO will be able to constrain the proton composition of ultrahigh-energy cosmic rays in scenarios that feature very strong source evolution and in which protons are accelerated to extremely high energies. Although gamma-ray observations are generally more sensitive to decaying particles than neutrino observations, PUEO is expected to set the strongest neutrino-detector constraints above $10^{19}$~eV. PUEO will also provide the strongest constraints on some models of cosmic strings.

\end{abstract}

\maketitle

\section{Introduction}

Neutrinos with energies above 1 EeV (1 EeV = $10^{18}$ eV) \cite{Ackermann:2022rqc} have not yet been observed. Such particles are expected to be produced through the interactions of ultrahigh-energy cosmic rays (UHECRs), both in their sources and during propagation. More specifically, the interactions of UHECRs with cosmic photon backgrounds are predicted to generate a flux of so-called ``cosmogenic'' neutrinos~\cite{Berezinsky_1969, Kotera:2010yn, Ahlers_2012, Batista_2019}, whose spectrum and intensity depends on the chemical composition of the UHECRs as well as on the redshift distribution of UHECR sources. Unlike cosmic rays, neutrinos are not deflected by magnetic fields or attenuated by cosmic photon backgrounds. Because of this, observations of ultrahigh-energy neutrinos could provide an essential means by which to identify the sources of UHECRs. Furthermore, ultrahigh-energy neutrinos can be used to probe new fundamental physics by observing particles at extremely high energies and over cosmological baselines. 

Many ultrahigh-energy neutrino experiments are currently  under construction or in development \cite{Ackermann:2022rqc}. These include, but are not limited to, in-ice and water detectors such as the Radio Neutrino Observatory in Greenland (RNO-G) \cite{RNO-G:2020rmc}, IceCube-Gen2 \cite{IceCube-Gen2:2021rkf},  KM3NeT \cite{KM3Net:2016zxf}, Baikal-GVD \cite{Baikal-GVD:2019kwy}, P-ONE \cite{P-ONE:2020ljt}, and TRIDENT \cite{TRIDENT:2022hql}; radio detectors operating from the upper atmosphere or in space such as PUEO \cite{Vieregg:2021nan}; space-based detectors such as POEMMA   \cite{Venters:2019xwi}; air-shower detectors on the ground such as  BEACON \cite{Zeolla:2023iqq}, GRAND \cite{Kotera:2021hbp}, TAMBO \cite{Thompson:2023pnl}, and Trinity \cite{2022icrc.confE1234W}; and moon-based detectors such as LUNASKA \cite{2011MNRAS.410..885J}. These experiments will provide unprecedented insights into the ultrahigh-energy universe, potentially shedding light on the sources of UHECRs, the nature of dark matter, and other new fundamental physics. The most imminent of these experiments is the Payload for Ultrahigh Energy Observations (PUEO). PUEO is a long-duration balloon-based experiment that launched in December 2025 and will complete a 30-day flight over Antarctica \cite{PUEO_talk}. PUEO is a successor to the Antarctic Impulse Transient Antenna (ANITA) program \cite{ANITA:2019wyx}, and will observe Askaryan radiation from ultrahigh-energy neutrinos from a high-altitude platform. With its 30-day flight, PUEO will provide the greatest sensitivity to date to neutrinos with energies in the range of $1 - 1000 \, {\rm EeV}$~\cite{PUEO_whitepaper}.

In this paper, we explore the scientific reach of PUEO, both to cosmogenic neutrinos and to selected scenarios involving exotic physics. In Sections~\ref{cosmogenic} and \ref{exotic}, we describe our calculations of the neutrino spectra predicted in various scenarios, and in Section~\ref{constraints} we present our forecast for the sensitivity of PUEO to these models. In Section~\ref{discussion}, we summarize our results and their implications.

\section{Cosmogenic neutrinos}
\label{cosmogenic}

The intensity and shape of the cosmogenic neutrino spectrum is highly sensitive to the chemical composition of UHECRs. This composition, however, is not yet well-known, especially at the highest energy, and there remains some debate regarding the composition inferred by the Pierre Auger Observatory and the Telescope Array \cite{Plotko_2023, PierreAuger:2024flk, PierreAuger:2023htc, PierreAuger:2022atd, TelescopeArray:2024buq, TelescopeArray:2020bfv, TelescopeArray:2018bep}. Furthermore, while it has often been assumed that the maximum energy of each cosmic-ray species is proportional to its atomic number, recent models have suggested that there could exist an additional pure-proton component among the highest-energy cosmic rays \cite{ Das:2020nvx, Ehlert_2024, Muzio2023_protons, Lu_Yuan_2025,KM3NeT_cosmogenic}. Because the magnetic deflections and mean free paths of UHECRs are highly dependent on their composition, a precise understanding of the UHECR composition is essential for identifying the sources of these extremely energetic particles.

During propagation, ultrahigh-energy protons interact with the cosmic microwave background (CMB) and the diffuse extragalactic background light (EBL), generating charged pions through the Greisen-Zatsepin-Kuzmin (GZK) process \cite{Greisen, Zatsepin_Kuzmin}. The subsequent decays of these pions produce ultrahigh-energy neutrinos. Cosmic-ray nuclei can also interact with the CMB  and EBL through the process of photodisintegration, but this leads to a flux of cosmogenic neutrinos that is suppressed relative to that predicted from protons~\cite{Anchordoqui_2007,Hooper:2004jc,Ave:2004uj,Allard:2006mv}. Because of this, the cosmogenic neutrino flux provides a direct observational link to the chemical composition of UHECRs. In particular, while an additional proton component at the highest energies might not appreciably affect the shape of the UHECR spectrum, it could significantly enhance the resulting flux of ultrahigh-energy neutrinos. A nondetection of ultrahigh-energy neutrinos, in turn, could be used to place an upper limit on the fraction of protons in the overall UHECR population. We parameterize this in terms of the fraction of UHECRs at $10^{19.55}$ eV that are protons, $f_p$, following the definitions in Ref.~\cite{van_Vliet_2019, IceCubeCollaborationSS:2025jbi}. In this section we assess the sensitivity of PUEO to this quantity. 

To calculate the spectrum of UHECR protons and cosmogenic neutrinos, we use the simulation framework CRPropa~3.2 \cite{crpropa3}. Following Ref~\cite{van_Vliet_2019}, we inject protons from an isotropic distribution of sources with a comoving injection rate density given by
\begin{equation}
\frac{dN}{dt}(z) \propto \begin{cases} 
(1+z)^m &  m \leq 0 \\
(1+z)^m & m > 0 \text{ and } z \leq 1.5\\
2.5^m & m > 0 \text{ and } z > 1.5,
\end{cases}
\end{equation}
up to a maximum redshift of $z = 4$. While the source evolution is not necessarily expected to follow a power law, this parameterization serves as a reasonable approximation for realistic source distributions. For reference, a source distribution that follows the star formation rate corresponds to $m \sim 3.4$ up to $z = 1$ and $m \sim -0.3$ at $z \ge 1$~\cite{PierreAuger:2023htc,Ehlert:2022jmy,Hopkins:2006bw}

For the spectrum of UHECR protons injected from their sources, we adopt a power-law parameterization with an exponential cutoff above a maximum energy, 
\begin{align}
\frac{dN}{dE} \propto E^{-\gamma} e^{-E/E^{\text{max}}_p}.
\end{align}
Our simulation accounts for energy losses due to interactions with the CMB and the EBL \cite{Gilmore_EBL} via electron pair production, photopion production, nuclear decay, and adiabatic losses due to the expansion of the universe. Note that the resulting cosmogenic neutrino flux depends on the average distance traveled by the protons (if the sources are farther away, more neutrino production will take place). Due to this, the proton fraction, $f_p$, and the source evolution index, $m$, are approximately degenerate parameters in the evaluation of the cosmogenic neutrino spectrum.

\begin{figure}[t]
    \centering
    \includegraphics[width = 1\linewidth]{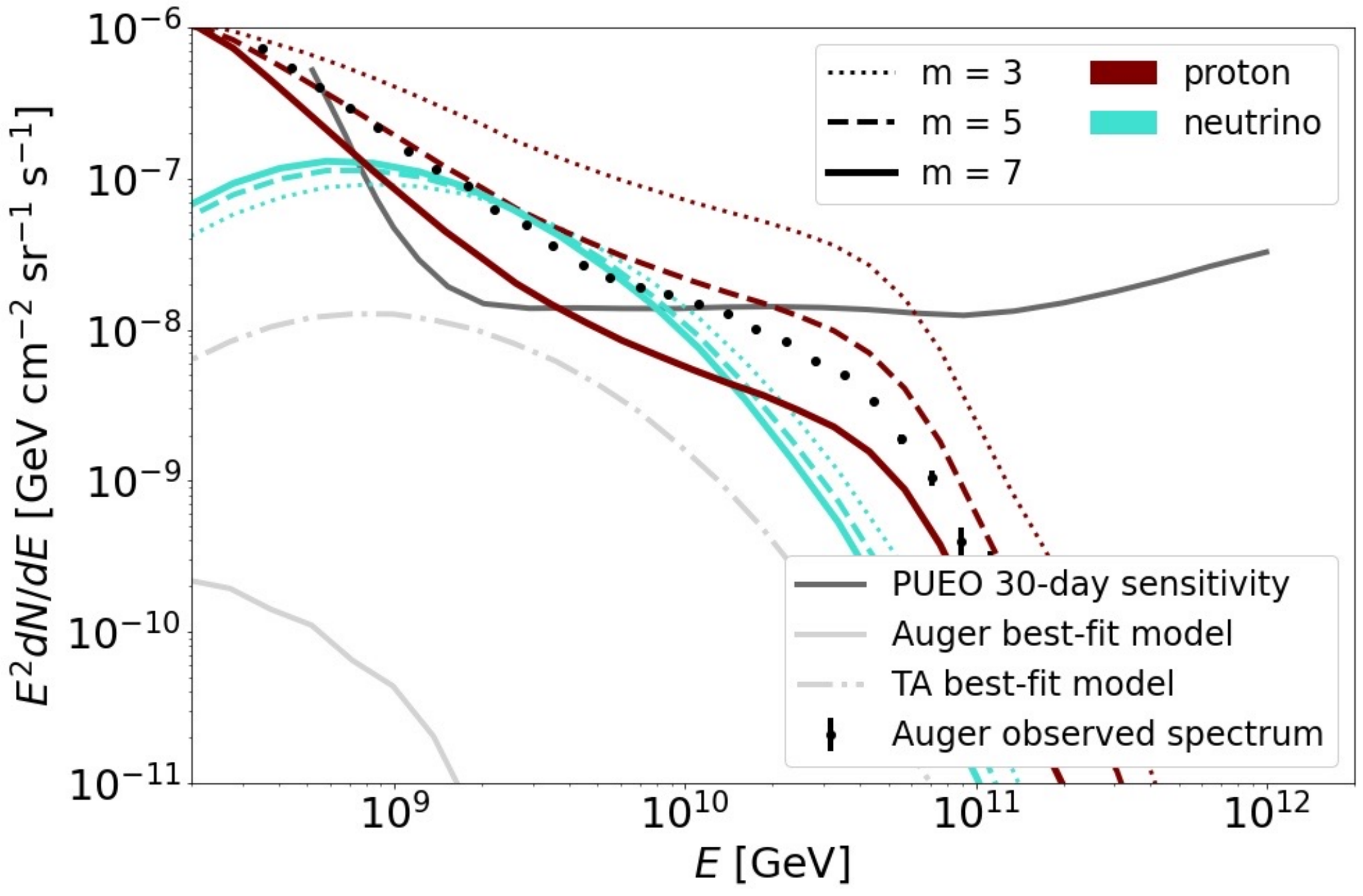}
    \caption{\label{cosmogenic_spectra} The spectra of UHECR protons (brown) and their corresponding cosmogenic neutrinos (teal). The colored spectra are normalized such that they produce (on average) 2.3 events in PUEO's upcoming flight. The varying linestyles correspond to different source evolution models ($m=3,5,7$). The PUEO 30-day sensitivity is shown by the solid dark gray line, while the black scatter points indicate the cosmic-ray spectrum as measured by the Pierre Auger Observatory \cite{Auger_spec_2017}. The light gray curves represent the neutrino spectra predicted from the Auger and Telescope Array best-fit cosmic-ray models. In this figure, we have adopted an injected spectrum with an index of $\gamma = 2.0$ and an energy cutoff of $E_{\text{max}} = 100$ EeV.}     
\end{figure}

If PUEO does not observe any ultrahigh-energy neutrinos during its flight, it will be able to place a  90\% confidence level upper limit on this flux at a level corresponding to an expected number of 2.3 events (based on Poisson statistics). To translate this into a projected upper limit on the proton fraction, $f_p$, we first calculate the cosmogenic neutrino spectrum for a variety of parameter choices. We then use the effective area from Ref.~\cite{Luszczak:20236b} to normalize the overall UHECR spectrum such that PUEO is predicted to observe a mean number of 2.3 events. Simulated proton and neutrino spectra are shown in Figure~\ref{cosmogenic_spectra} for three choices of the source evolution index. For the cases $m=3$ and $m=5$, PUEO is predicted to observe 2.3 or more events only if the normalization of the ultrahigh-energy proton spectrum is (unrealistically) set to exceed the overall cosmic-ray spectrum measured by Auger~\cite{Auger_spec_2017}. For the somewhat extreme case of $m=7$, however, PUEO could expect to detect cosmogenic neutrinos if a significant fraction of the highest-energy cosmic rays are protons. In Figure \ref{cosmogenic_spectra} we have adopted an injected spectral index of $\gamma=2.0$ and a maximum energy of $E^{\text{max}}_p$ = 100 EeV. When we later evaluate the projected constraints on $f_p$ from PUEO, we will consider a range of values for $\gamma$ and $E^{\text{max}}_p$. For reference, we also show in this figure the neutrino spectra that are predicted in the scenarios which provide the best fit to the data of the Pierre Auger Observatory \cite{Auger_best_fit} and the Telescope Array \cite{TA_best_fit}. 
The best fit from the Pierre Auger Collaboration assumes flat source evolution ($m = 0$) and corresponds to $\gamma=0.96$, $E^{\rm max}_p = 4.8 \, {\rm EeV}$, and source injection fractions of 0\%, 67.3\%, 28.1\%, 4.6\%, and 0\% for H, He, N, Si, and Fe, respectively. The Telescope Array Collaboration assumes $m=3$ and finds a best fit for $\gamma=2.06$, $E^{\rm max}_p = 182 \, {\rm EeV}$, and source injection fractions of 0\%, 99.2\%, 0\%, 0\%, and 0.8\% for H, He, N, Si, and Fe.

\section{Exotic physics}
\label{exotic}

In addition to the interactions of UHECRs, ultrahigh-energy neutrinos are predicted to be produced in a variety of exotic physics models. These scenarios include the decay of superheavy dark matter particles~\cite{Murase:2012xs,Guepin:2021ljb,Berghaus:2025jwb}, the evaporation of primordial black holes~\cite{Airoldi:2025opo,Boccia:2025hpm,Singh:2025iyb,Anchordoqui:2025xug}, and radiation from cosmic strings or other topological defects~\cite{Berezinsky:2011cp,Lunardini:2012ct,Creque-Sarbinowski:2022mex}. Here we consider two classes of exotic scenarios for ultrahigh-energy neutrino production: decay of superheavy dark matter and the radiation of moduli from cosmic string cusps. 

\subsection{Neutrinos From Superheavy Dark Matter Decay}

Approximately 85\% of matter in the universe is ``dark'', i.e. non-electromagnetically interacting. The existence of dark matter is inferred from the dynamics of galaxies and galaxy clusters, the large scale structure of the universe, and the cosmic microwave background, all of which indicate that the majority of the matter density is dark~\cite{Bertone:2016nfn}. A wide variety of dark matter candidates have been proposed, ranging from ultralight particles such as axions \cite{axions}, to massive compact objects such as primordial black holes \cite{PBH}. PUEO could be sensitive to dark matter in the form of unstable ultraheavy ($m_{\rm DM} \gsim 10^{8} \, {\rm GeV}$) particles. Such particles could have been produced in the early universe through gravitational production~\cite{SH_WIMP,Kolb:2023ydq}, thermal freeze-in~\cite{Kolb:2017jvz}, or other mechanisms~\cite{Babichev:2018mtd,Kim:2019udq,Dudas:2020sbq}.
The decays of these particles could result in the production of ultrahigh-energy Standard Model particles, including neutrinos. 
In this section, we consider the decay of a generic superheavy dark matter (SHDM) particle into ultrahigh-energy neutrinos and estimate the ability of PUEO to place constraints on such a decay channel.

For our calculation of the neutrino spectrum from SHDM decay, we follow the approach of Ref.~\cite{fiorillo2023} and utilize the 
publicly-available code \texttt{HDMSpectra} \cite{HDMSpectra}. We focus on the case of a neutrinophilic decay channel, $\chi \rightarrow \nu\overline{\nu}$, and assume equal branching ratios to each neutrino flavor.

To determine the spectrum of ultrahigh-energy neutrinos from SHDM decay, we integrate over the Galactic and extragalactic dark matter distributions. The sky-averaged neutrino flux from decays in the Galactic halo is given by
\begin{eqnarray}
\frac{d\Phi_{\nu}^{\text{Gal}}}{dE_{\nu}} = \frac{1}{4\pi} \frac{dN_{\nu}}{dE_{\nu}}\int \frac{\rho_{\text{DM}}(s, b, l)}{4\pi\tau_{\text{DM}}m_{\text{DM}}}\sin(b)dldbds, 
\end{eqnarray}
where $\rho_{\text{DM}}(s, b, l)$ is the Galactic dark matter distribution, $l$ and $b$ are Galactic longitude and latitude, respectively, and $s$ is the distance from Earth. $m_{\rm DM}$ and $\tau_{\rm DM}$ are the mass and lifetime of the SHDM particle, respectively, and $dN_{\nu}/dE_{\nu}$ is the spectrum of neutrinos produced per decay. For the Galactic dark matter distribution we adopt a Navarro-Frenk-White (NFW) profile following Ref.~\cite{NFW}, which is parameterized by 
\begin{equation}
\rho_{\text{DM}} \propto \frac{1}{\bigg(\dfrac{r(s, b, l)}{r_s}\bigg)\bigg(1 + \dfrac{r(s, b, l)}{r_s}\bigg)^2},
\end{equation}
where  $r(s, b, l) = \sqrt{s^2 + R_{\rm GC}^2 - 2sR_{\rm GC}\cos{b}\cos{l}}$ is the distance from the Galactic Center. We adopt $r_s = 20$ kpc as the scale radius of the halo, and $R_{\rm GC}=8.25 \, {\rm kpc}$ \cite{McMillan:2016jtx}. We normalize this density profile such that $\rho_{\rm DM} = 0.4$ GeV/cm$^3$ at $r=R_{\rm GC}$ \cite{Read:2014qva}.

\begin{figure}[t]
    \centering
    \includegraphics[width = 1\linewidth]{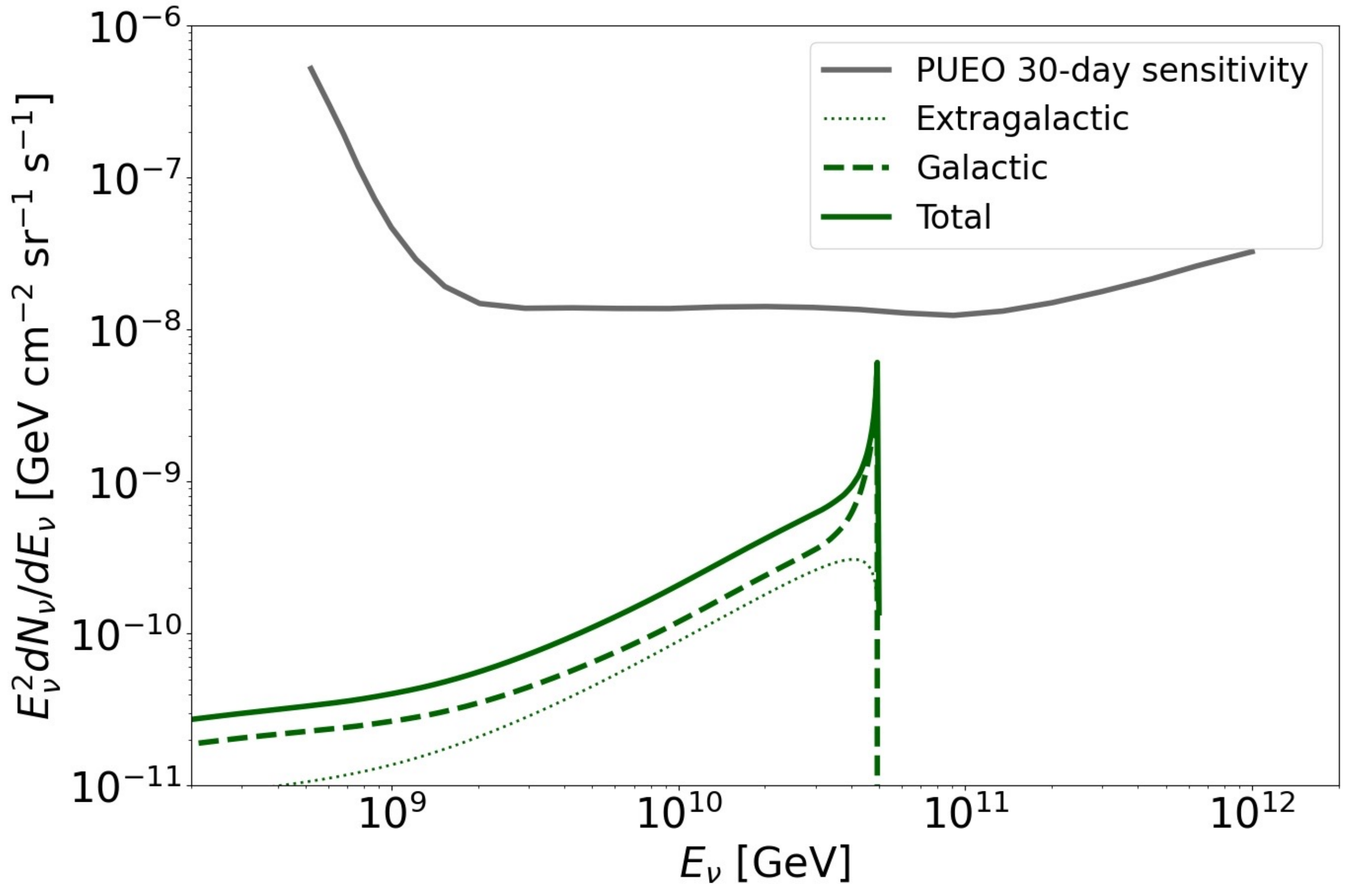}
    \caption{\label{shdm_spectra} The spectrum of ultrahigh-energy neutrinos from the decay of superheavy dark matter particles with $m_{\text{DM}} = 10^{11}$ GeV and $\tau_{\rm DM} = 10^{30}$ s. This value of the lifetime approximately saturates the current constraints from gamma-ray observations~\cite{Das:2023wtk}. The solid line indicates the total neutrino spectrum, while the dashed and dotted lines show the contribution from the Galactic and cosmological dark matter distributions, respectively. The PUEO 30-day sensitivity is shown by the dark gray curve.}     
\end{figure}

Similarly, the spectrum of ultrahigh-energy neutrinos produced from extragalactic dark matter decay is given by integrating over the cosmological distribution,    
\begin{equation}
\frac{d\Phi_{\nu}^{\text{EG}}}{dE_{\nu}} = \frac{\Omega_{\text{DM}}\rho_c}{4\pi\tau_{\text{DM}}m_{\text{DM}}}\int \frac{dz}{H(z)}\frac{dN_{\nu}}{dE_{\nu}} \bigg\vert_{E_\nu(1+z)},
\end{equation}
where $\Omega_{\text{DM}} = 0.265$ is the fraction of universe's energy density in dark matter, $\rho_c = 4.79\times10^{-6}$ GeV/cm$^3$ is the critical density, and $H(z) = H_0\sqrt{\Omega_{\Lambda} + \Omega_\text{M}(1+z)^3}$ is the Hubble rate. We further adopt $\Omega_{\Lambda} = 0.685$, $\Omega_\text{M} = 0.315$, and $H_0=67.4 \, {\rm km/s/Mpc}$~\cite{Planck:2018vyg}.

In Figure \ref{shdm_spectra}, we show the neutrino spectrum from SHDM decays for $m_{\text{DM}} = 10^{11}$ GeV and for a lifetime of $\tau_{\rm DM} = 10^{30} \, {\rm s}$. This value of the lifetime approximately saturates the current constraints from gamma-ray observations~\cite{Blanco:2018esa,Das:2023wtk}. The total neutrino flux is represented by a solid line, while the dashed and dotted curves show the contributions from the Galactic and cosmological distributions, respectively. The sharp peak in the Galactic curve is due to the contribution of neutrinos at $E_\nu = m_{\rm DM}/2$, while in the cosmological case this peak is smoothed out by the integration over redshift. The PUEO 30-day sensitivity is represented by the dark gray curve. For the case shown, PUEO is predicted to observe a mean number of 0.02 events over its 30-day flight (see Table~\ref{Nexp}).

\subsection{Neutrinos From Cosmic Strings}

In some models, phase transitions in the early universe can result in the formation of topologically stable configurations of gauge and Higgs fields, known as topological defects~\cite{Kirzhnits:1972ut,Kibble:1976sj,Preskill:1984gd,Vilenkin:1984ib}. In particular, the spontaneous breaking of a $U(1)$ symmetry can result in one-dimensional defects known as cosmic strings~\cite{Nielsen:1973cs}. 

Over the course of their evolution, cosmic string loops can emit a potentially observable flux of ultrahigh-energy neutrinos. While cosmic strings form as smooth objects, they eventually cross, self-cross, and reconnect, leading them to emit radiation at high or ultrahigh energies. The spectrum of this radiation depends on many model-dependent factors, including the length of the cosmic string loops. The loops also develop localized features, ``kinks" and ``cusps," where the radiation efficiency can be greatly enhanced. An observable neutrino flux could be achieved through the emission from the strings of weakly-coupled massive scalar particles, known as moduli, which decay into hadrons. In particular, if the coupling of moduli to strings is stronger than that of gravity, moduli emission will be the primary form of cosmic string radiation. Ultrahigh-energy neutrino emission from cosmic string cusps is discussed in Ref.~\cite{Berezinsky:2011cp}, and from cosmic string kinks in Ref.~\cite{Lunardini_2012}.

As an example, we consider here the neutrino spectrum predicted from cosmic string cusps as calculated in Ref.~\cite{Berezinsky:2011cp} (the spectrum from cosmic string kinks is similar~\cite{Lunardini_2012}). 
This spectrum depends on several free parameters, including the coupling of moduli to the cosmic strings, $\alpha$, and the mass of the moduli, $m$.
For $\alpha \sim 1$, the modulus coupling to Standard Model particles will be strongly suppressed, so we focus on the parameter space in which $\alpha$ is very large (values as large as $\alpha \sim 10^{15}$ have been considered in Ref.~\cite{Goldberger:1999un}, for example).

\begin{figure}[t]
    \centering
    \includegraphics[width = 1\linewidth]{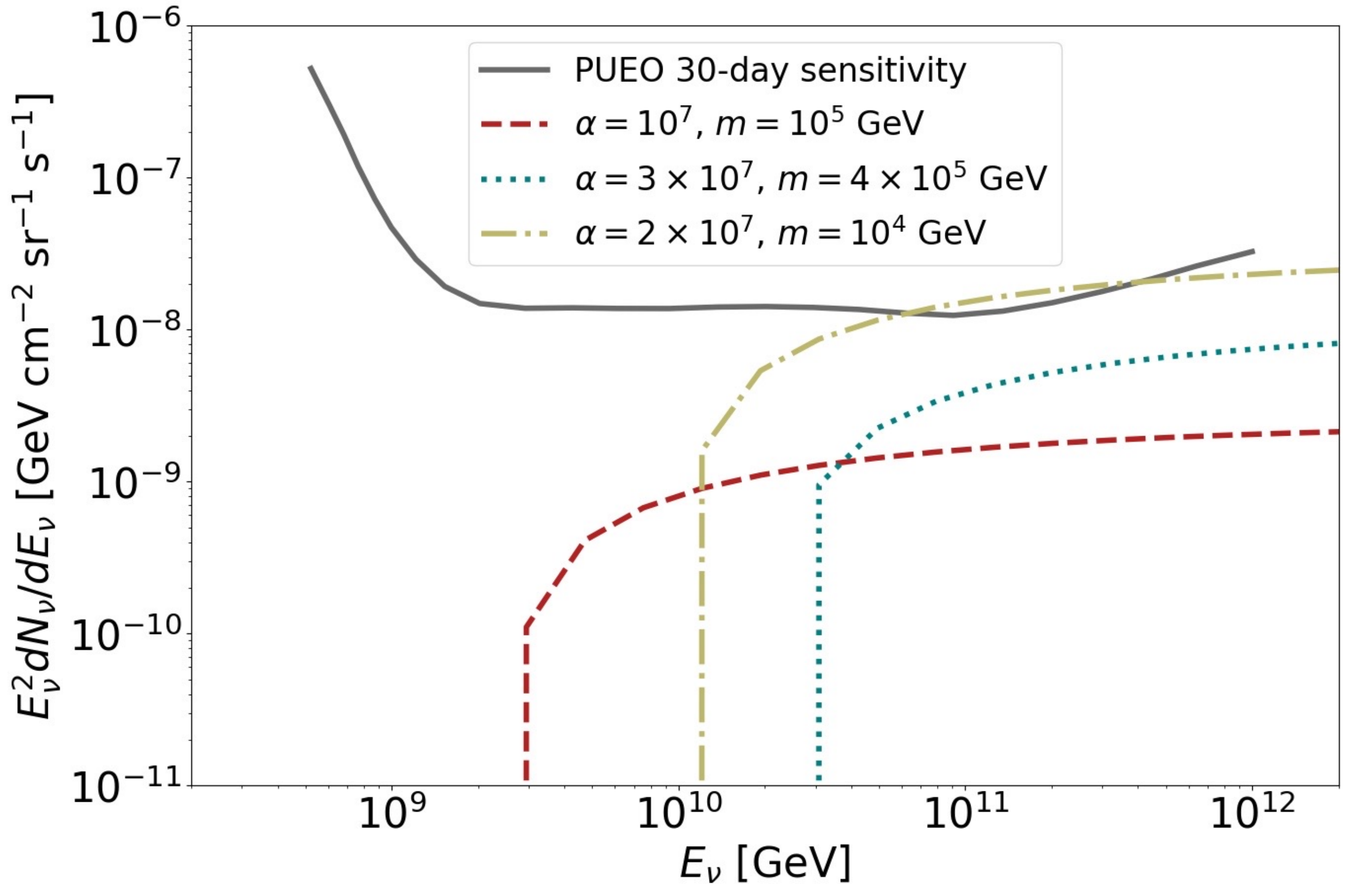}
    \caption{\label{cosmic_strings} The spectra of ultrahigh-energy neutrinos from cosmic string radiation, as calculated in the model of Ref.~\cite{Berezinsky:2011cp}. The three curves correspond to different values of the coupling strength of moduli to cosmic strings $\alpha$, and the modulus mass $m$. The PUEO 30-day sensitivity is shown by the gray curve. }     
\end{figure}

The spectrum of neutrinos radiated from cosmic strings is given by 
\begin{align}
E^2\frac{dN}{dE} &= 2.5\times 10^{-9} \, {\rm GeV} \, {\rm cm}^{-2} \, {\rm s}^{-1} \, {\rm sr}^{-1} \\
&\times \bigg(\frac{1}{p}\bigg) \bigg(\frac{\alpha}{10^7}\bigg)^2 
\bigg(\frac{10^5 \, {\rm GeV}}{m}\bigg)^{1/2} \bigg(\frac{z_{\nu}}{200}\bigg)^{1/2} \nonumber \\
&\times \bigg(1 -\sqrt{\frac{1 + z_{\rm min}(E_{\nu})}{1+z_\nu}}\bigg), \nonumber 
\end{align}
where $p$ is the reconnection probability for cosmic strings, and $z_{\nu}$ is the neutrino horizon. The minimum redshift from which a neutrino radiated from a cosmic string could plausibly be observed is determined by the rate of bursts and by the minimum radiated neutrino energy, corresponding to the smallest surviving loop size at a given redshift. We take the minimum redshift in our integration to be
\begin{align}
1 + z_{\rm min}(E_{\nu}) = \bigg(\frac{E_0}{E_{\nu}}\bigg)^{4/7} \bigg(\frac{\alpha}{10^7}\bigg)^{8/7}\bigg(\frac{z_*}{z_\nu}\bigg)^{3/7},
\end{align}
where $z_*$ is the redshift at which cosmic string loops transition from modulus-dominated radiation to gravity-dominated radiation. In our calculations, we take $p=1$, $z_{\nu}=z_*= 200$, and $E_0 =  2.7 \times 10^{13} \, {\rm GeV}$. We show our results in Figure \ref{cosmic_strings} for a few values of $\alpha$ and $m$.

\section{Projections and prospects}
\label{constraints}

In Table~\ref{Nexp}, we show the number of neutrinos predicted to be detected by PUEO in each of the models we have considered in this study. The number of such expected events is calculated as~\cite{PUEO_whitepaper}
\begin{equation}
N_{\text{exp}} = T_{\text{obs}} \int \frac{dN_\nu}{dE_\nu}A\Omega_{\text{eff}}dE,
\end{equation}
where $T_{\text{obs}} = 30$ days, $A\Omega_{\text{eff}}$ is PUEO's all-flavor effective area~\cite{Luszczak:20236b}, and $dN_\nu/dE_\nu$ is neutrino spectrum. Note that the expected number of neutrino events for the Telescope Array best-fit model is lower here than the value of 0.9 shown in Ref.~\cite{PUEO_whitepaper}; this is due to the fact that we do not adopt a minimum source redshift and use a more recent fit to the Telescope Array data. 

\begin{table}[t]
    \renewcommand{\arraystretch}{1.2}
    \begin{ruledtabular}
    \begin{tabular}{cc}
    \textbf{Cosmogenic} & $N_{\text{exp}}$ \\
    \hline
    Auger Best-Fit Model & 0.0003 \\
    Telescope Array Best-Fit Model & 0.37\\
    \hline
    \textbf{Dark Matter Decay} & $N_{\text{exp}}$ \\
    \hline
    $m_{\rm DM} = 10^{11} \, {\rm GeV}, \tau_{\rm DM} = 10^{30} \, {\rm s}$ & 0.02 \\
    \hline
    \textbf{Cosmic Strings} & $N_{\text{exp}}$ \\
    \hline
    $\alpha = 10^7, m = 10^5$ GeV & 0.13 \\
    $\alpha = 3\times 10^7, m = 4\times 10^5$ GeV & 0.25 \\
    $\alpha = 2\times 10^7, m = 10^4$ GeV & 1.1 \\
    \end{tabular} 
    \end{ruledtabular}
    \caption{The number of neutrinos predicted to be detected by PUEO in each of the models considered in this study.}
    \label{Nexp}
\end{table}

\begin{figure}[t]
    \centering
    \includegraphics[width = 1\linewidth]{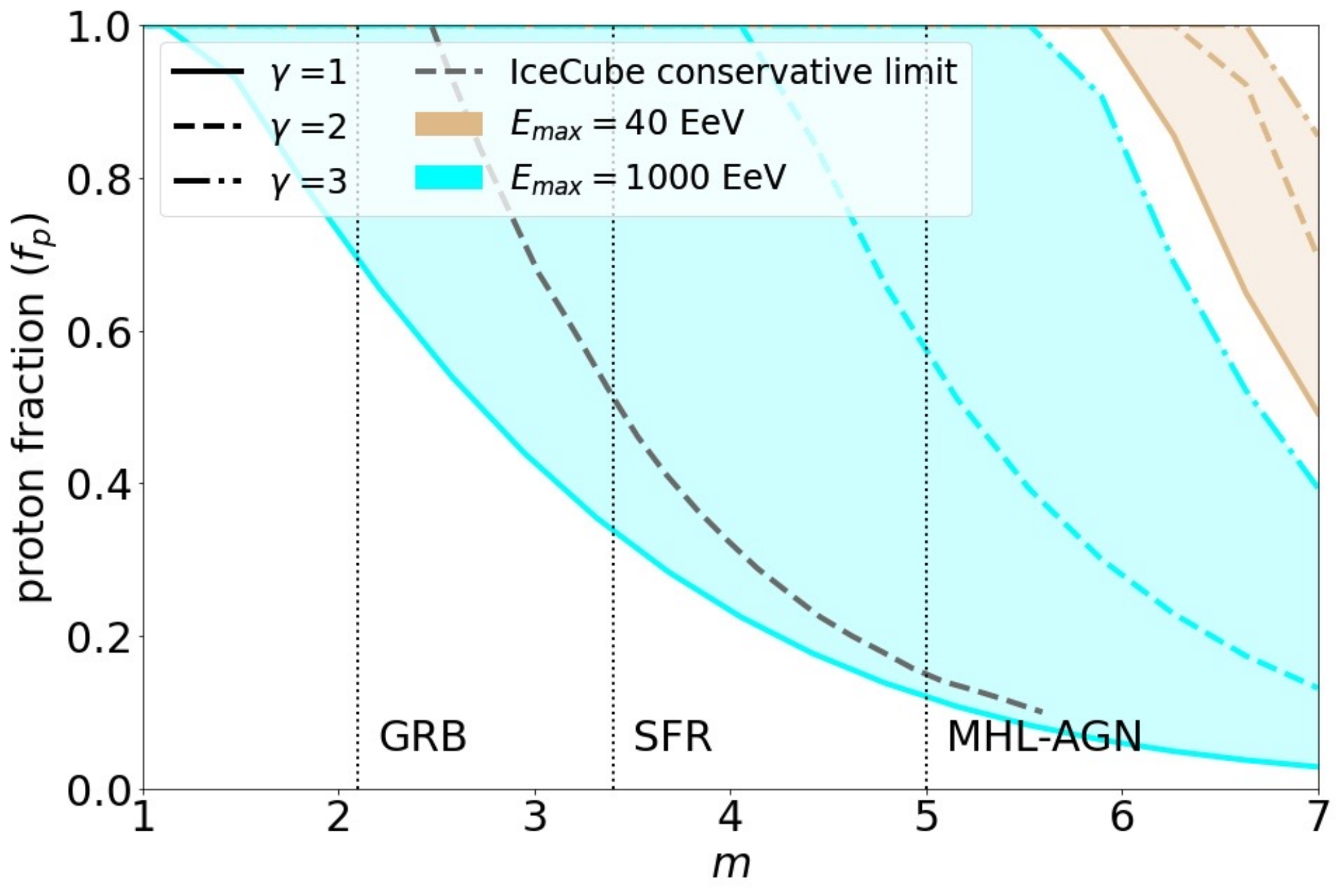}
    \caption{\label{proton_fraction} 
The regions of the UHECR parameter space that would be ruled out by PUEO if it does not detect any ultrahigh-energy neutrino events in its 30-day flight. Such a nondetection would exclude some models with both very strong source evolution, $m$, and a large proton fraction, $f_p$. Results are shown for $\gamma=1.0$, 2.0, and 3.0 and $E_{\rm max}=40$ and $1000 \, {\rm EeV}$. The proton fraction is defined at $10^{19.55}$ eV. The dashed gray line shows the most conservative limit placed by Ref.~\cite{IceCubeCollaborationSS:2025jbi}. The vertical lines indicate the source evolution of gamma-ray bursts (GRB), sources that follow the star formation rate (SFR), and medium-high-luminosity active galactic nuclei (MHL-AGN).}     
\end{figure}

In Figure \ref{proton_fraction}, we show the regions of the UHECR parameter space that would be ruled out by PUEO if it does not detect any events in its 30-day flight. In particular, such a nondetection would exclude some models with both very strong source evolution and a large proton fraction, $f_p$. We show results for $\gamma=1.0$, 2.0, and 3.0 and $E_{\rm max}=40$ and $1000 \, {\rm EeV}$. The vertical lines indicate the source evolution of a few representative UHECR source candidates: gamma-ray bursts (GRB)\footnote{It has not been concluded whether the GRB rate follows an SFR distribution.}\cite{Wanderman:2009es}, sources that follow the star formation rate (SFR) \cite{Yuksel:2008cu}, and medium-high-luminosity active galactic nuclei (MHL-AGN) \cite{Hasinger:2005sb}. The dashed gray line shows the most conservative analogous limit from IceCube \cite{IceCubeCollaborationSS:2025jbi}. 

\begin{figure}[t]
    \centering
    \includegraphics[width = 1\linewidth]{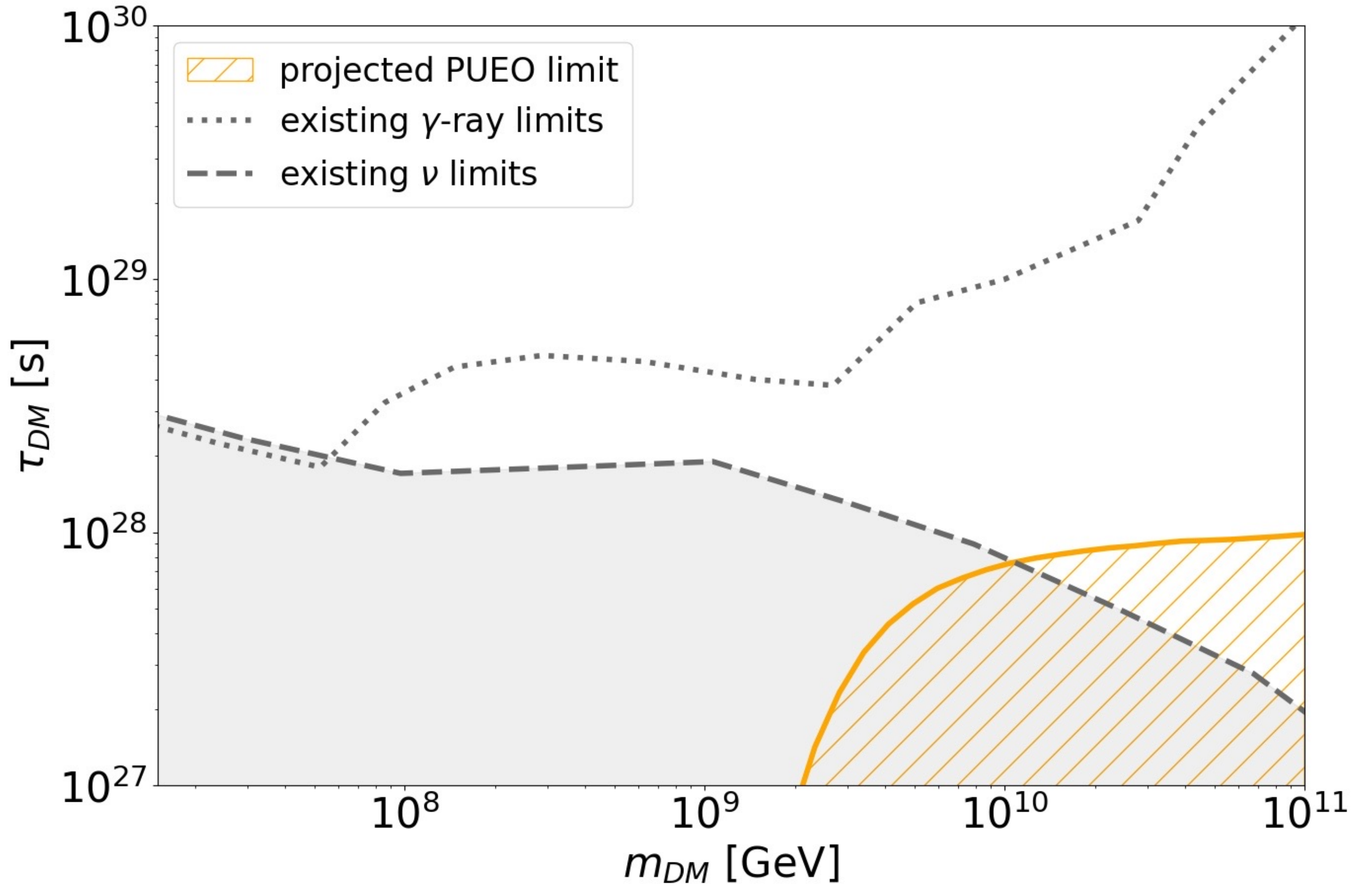}
    \caption{\label{shdm_limits} The projected limit from PUEO (assuming no events are observed) on the lifetime of superheavy dark matter (orange hatched). The shaded region indicates the parameter space that is already disfavored by existing neutrino observations, while the dashed line denotes the parameter space that is disfavored by existing gamma-ray observations. For $m_{\text{DM}} \gtrsim 10^{10}$ GeV, PUEO is expected to place the strongest neutrino-based limits on SHDM, although gamma-ray constraints will remain more restrictive.}   
\end{figure}

In Figure \ref{shdm_limits}, we show as an orange hatched region the projected limit from PUEO (assuming no events are observed) on the lifetime of SHDM. The shaded region indicates the parameter space that is already disfavored by existing neutrino observations, while the dashed line denotes the parameter space that is disfavored by existing gamma-ray observations~\cite{Arguelles_DMdecay}. For $m_{\text{DM}} \gtrsim 10^{10}$ GeV, PUEO is expected to place the strongest neutrino-based limits on SHDM, although gamma-ray constraints will remain more restrictive. 

Because the assumptions we use for radiation by cosmic strings are quite model-specific, we do not attempt to place limits on the broader cosmic strings parameter space. Instead, in Table \ref{Nexp} we evaluate the number of detectable neutrinos for three benchmark sets of cosmic string parameters outlined in Ref. \cite{Berezinsky:2011cp}.

\section{Summary and Discussion}
\label{discussion}

The Payload for Ultrahigh Energy Observation (PUEO) is a long-duration balloon-based experiment that will provide leading sensitivity to ultrahigh-energy neutrinos with energies in the range of $1 - 1000$ EeV. In this work, we explored the physics and astrophysics that will be probed by this experiment. In particular, we focused on PUEO's ability to constrain the proton composition of the UHECR spectrum and the lifetime of superheavy dark matter particles, and to search for ultrahigh-energy neutrinos from cosmic strings.  
 
During propagation, UHECR protons interact with the CMB and EBL to produce pions through the GZK process. These pions decay into gamma-rays and neutrinos, thus providing a direct link between UHECR protons and cosmogenic neutrinos. Using CRPropa~3.2, we generated spectra of cosmogenic neutrinos and evaluated the ability of PUEO to place constraints on the proton fraction of UHECRs.
We find that PUEO will be able to constrain the proton fraction in scenarios which feature strong source evolution and in which protons are accelerated to extremely high energies. 

In addition to cosmogenic neutrinos, ultrahigh-energy neutrinos could also be produced through the decays of superheavy dark matter (SHDM) particles. In our analysis, we considered SHDM decay to neutrinos, $\chi \rightarrow \nu\overline{\nu}$, and generated simulated spectra using the publicly-available software \texttt{HDMSpectra}. To infer the expected neutrino flux at Earth, we integrated over the Galactic and extragalactic dark matter distributions and compared the results to the PUEO sensitivity. We find that while PUEO will provide the greatest sensitivity to UHE neutrinos from SHDM, such scenarios will remain more strongly constrained by gamma-ray observations.

Finally, we also considered ultrahigh-energy neutrinos produced by cosmic strings. A variety of particle physics models predict cosmic string loops which can radiate energy in the form of moduli, weakly-interacting massive scalar particles that produce neutrinos in their decays. 
We compared the PUEO sensitivity to the neutrino spectra predicted from cosmic string cusps as described in Ref.~\cite{Berezinsky:2011cp}. We find that for some values of the modulus coupling and mass, PUEO could plausibly detect neutrinos from such objects. 

In this work, we have not considered ultrahigh-energy neutrinos from proposed accelerators of UHECRs, such as  active galactic nuclei~\cite{1991PhRvL..66.2697S},  gamma-ray burst afterglows~\cite{Waxman:1999ai}, galaxy clusters~\cite{Murase_2008_GalaxyClusters}, or newborn pulsars~\cite{Fang:2013vla}  (see also Ref.~\cite{Fang:2025wpr} and references therein). If pion production occurs efficiently in these sources, this could lead to potentially observable fluxes of ultrahigh-energy neutrinos. Ref.~\cite{PUEO_whitepaper} estimates the ultrahigh-energy neutrino fluxes for a few source classes.

Over the next decade, upcoming neutrino observatories such as IceCube-Gen2, GRAND, and POEMMA will probe even deeper into the $\gtrsim 1$ EeV regime of neutrino astrophysics. This includes the energy range over which cosmogenic neutrinos are expected, but also encompasses higher energies at which neutrinos could only be produced through top-down mechanisms, such as superheavy dark matter decay or radiation from cosmic strings. In this work, we have demonstrated the exciting potential of upcoming ultrahigh-energy neutrino observatories to inform our understanding of the source evolution and composition of the UHECRs and to probe physics beyond the Standard Model at the highest observed energies. 

\nocite{Sherman_PUEO_2025}

\smallskip

\begin{acknowledgments}
K.F. acknowledges support from the National Science Foundation (PHY-2238916, PHY-2514194) and the Sloan Research Fellowship. This work was supported by a grant from the Simons Foundation (00001470, KF).  D.H. is supported by the Office of the Vice Chancellor for Research at the University of Wisconsin-Madison, with funding from the Wisconsin Alumni Research Foundation.
\end{acknowledgments}

%

\end{document}